# Strongly Secure Communications Over the Two-Way Wiretap Channel

Alexandre J. Pierrot, *Student Member, IEEE* and Matthieu R. Bloch, *Member, IEEE*

*Abstract*—We consider the problem of secure communications over the two-way wiretap channel under a strong secrecy criterion. We improve existing results by developing an achievable region based on strategies that exploit both the interference at the eavesdropper's terminal and cooperation between legitimate users. We leverage the notion of channel resolvability for the multiple-access channel to analyze cooperative jamming and we show that the artificial noise created by cooperative jamming induces a source of common randomness that can be used for secret-key agreement. We illustrate the gain provided by this coding technique in the case of the Gaussian two-way wiretap channel, and we show significant improvements for some channel configurations.

## I. INTRODUCTION

IN CRYPTOGRAPHY, noise is often viewed as an impairment that should be handled at the physical layer before any encryption is applied. In contrast, the works of Wyner [1] and Csiszár and Körner [2] on the wiretap channel, and Maurer [3] and Ahlswede and Csiszár [4] on secret-key agreement from common randomness support the idea that there is much security to be gained by exploiting the noise present at the physical layer. This concept, colloquially known as *physical-layer security*, has attracted a growing interest and has led to numerous extensions of the wiretap channel model [5], [6].

In particular, the analysis of multi-user channel models with secrecy constraints has highlighted the pivotal role of cooperation, jamming, and feedback as means to increase secure communication rates. The two-way wiretap channel, in which users communicate over a noisy bidirectional channel while an eavesdropper observes interfering signals, combines all the aforementioned effects because users have the possibility of cooperating while simultaneously jamming the eavesdropper. This model was first investigated by Tekin and Yener [7], [8] who showed that jamming with noise or controlled interference between codewords could provide secrecy gains. However, this strategy called *cooperative jamming* does not exploit feedback. It was later shown by He and Yener [9] and Bloch [10] that strategies based on feedback can perform strictly better than cooperative jamming alone. Recently, El Gamal *et al.* [11] proposed an achievable region for the two-way wiretap channel combining cooperative jamming and a secret-key exchange mechanism to transfer secure rate between users.

From a practical perspective, the two-way wiretap channel captures some of the limitations of real systems because all communications are intrinsically rate-limited. We note that the two-way wiretap channel also generalizes many models; for instance, the works of Amariucai and Wei [12], Gündüz *et al.* [13], and Lai *et al.* [14] are special cases that focus on secure communication for one user only. Similarly, the model of Ardestanizadeh *et al.* [15] is a two-way wiretap channel in which one of the link is confidential and unheard by the eavesdropper. Many works on secret-key agreement with rate-limited public communication can be analyzed within this framework [16], [17] as well.

In this paper, we extend existing results in several directions.
We show that it is possible to design powerful coding schemes by partially decoupling the feedback and the interference and by relying on the following strategies:

- *Cooperative jamming*, which leverages codeword interference to give legitimate users an advantage over the eavesdropper by reducing its signal-to-noise ratio.
- *Key exchange*, by which one user sacrifices part of its secret rate to transmit a key to the other, which uses the key to encrypt a message, thus augmenting its own secret rate. Note that key exchange is a basic feedback mechanism that transfers secure rate from one user to the other, but does not generate new secrecy.
- *Key generation*, by which the users exploit the intrinsic randomness of the channel to generate a secret key. Since channels are noisy, the communication required to distill the key is inherently rate-limited. The key is then used by one of the users to encrypt a message, and to increase its secret communication rate.

We establish strong secrecy results, which require the eavesdropper to obtain a negligible amount of information instead of a negligible rate of information, by exploiting the concept of channel resolvability [18]–[22] to analyze cooperative jamming. Channel resolvability provides a conceptually convenient interpretation of cooperative jamming, which allows us to analyze what happens when transmitting beyond the capacity of the eavesdropper's channel.

The outline of this paper is as follows. In Section II, we provide definitions of a two-way wiretap channel and a wiretap code. In Section III, we derive a region of strongly secure rates achievable with cooperative jamming based on channel resolvability. This first step yields a result similar to what Tekin and Yener [7,

Theorem 2] obtained for weak secrecy. In Section IV, the region is improved by introducing the key exchange mechanism proposed in [9], [11]. We further extend this region by performing secret key generation from a source induced by the noise used in cooperative jamming. Finally, in Section V, we illustrate the achievable region in the Gaussian case. Section VI concludes the paper by discussing possible extensions of this work.

## II. PROBLEM STATEMENT

*Notation:* Consider two random variables $X$ and $Y$ taking values in alphabets $\mathcal{X}$ and $\mathcal{Y}$. Sample values of $X$ and $Y$ are denoted by $x$ and $y$, respectively; the joint probability law is denoted by $p_{XY}$, and the marginal probabilities are denoted by $p_X$ and $p_Y$, respectively. We suppose all probability laws either discrete or Gaussian. The *mutual information* between $X$ and $Y$ is the random variable:

$$I(X;Y) \triangleq i_{X;Y}(X;Y) \quad \text{where:} \quad i_{X;Y} \triangleq \log_2 \frac{p_{XY}}{p_X p_Y}.$$

We refer to the average of the mutual information random variable as the *average mutual information*, and denote it by

$$\mathbb{I}(X;Y) \triangleq \mathbb{E}(I(X;Y)),$$

the *entropy* corresponds to $\mathbb{H}(X) = \mathbb{I}(X;X)$. The weakly typical set with respect to distribution $p_X$ and $\epsilon$ is denoted $\mathcal{A}_\epsilon^n(X)$. Finally, we let $\delta(\epsilon)$ (resp. $\delta(n)$) be a short-hand notation for any function of $\epsilon$ (resp. $n$) that goes to zero as $\epsilon$ goes to zero (resp. $n$ goes to infinity). We also introduce the integer interval $[\![a,b]\!]$ consisting of all integers between $\lfloor a \rfloor$ and $\lceil b \rceil$.

We consider the problem of secure communication over a two-way wiretap channel illustrated in Fig. 1, in which:

- a legitimate user called Alice (or transmitter 1) sends message $M_1$ and estimates $M_2$;
- another legitimate user called Bob (or transmitter 2) sends message $M_2$ and estimates $M_1$;
- an eavesdropper called Eve observes $Z^n$.

We assume the channel is full-duplex, which means Alice and Bob communicate *simultaneously* over the channel. This assumption is relevant for some communication systems; however, it may be hard to realize in practice and many experimental communication systems operate with half-duplex, potentially yielding lower rates. The analysis of half-duplex channels is left as future work.

*Definition 1:* A *two-way wiretap channel* denoted by $\left(\mathcal{X}_1, \mathcal{X}_2, \mathcal{Y}_1, \mathcal{Y}_2, \mathcal{Z}, \{p_{Y_1^n Y_2^n Z^n | X_1^n X_2^n}\}_{n \geqslant 1}\right)$ consists of two arbitrary input alphabets $\mathcal{X}_1$ and $\mathcal{X}_2$, three arbitrary output alphabets $\mathcal{Y}_1$, $\mathcal{Y}_2$ and $\mathcal{Z}$, and a sequence of transition probabilities $\{p_{Y_1^n Y_2^n Z^n | X_1^n X_2^n}\}_{n \geqslant 1}$ such that:

$$\forall n \in \mathbb{N}^*, \forall (x_1^n, x_2^n) \in \mathcal{X}_1^n \times \mathcal{X}_2^n,$$
$$\sum_{y_1^n \in \mathcal{Y}_1^n} \sum_{y_2^n \in \mathcal{Y}_2^n} \sum_{z^n \in \mathcal{Z}^n} p_{Y_1^n Y_2^n Z^n | X_1^n X_2^n}(y_1^n, y_2^n, z^n | x_1^n, x_2^n) = 1. \quad (1)$$

We consider hereafter a memoryless wiretap channel, but we note that our approach and our results generalize in part to arbitrary channels using information spectrum methods [20].

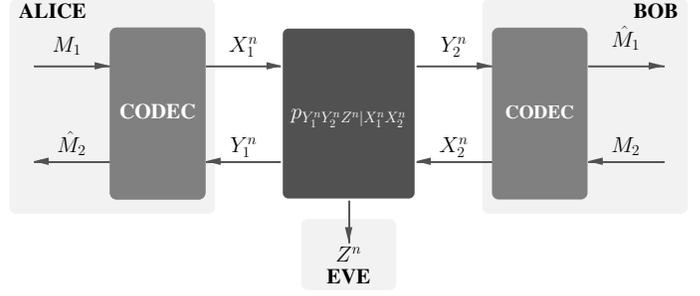

Fig. 1. Communication over a two-way wiretap channel

*Definition 2:* A memoryless two-way wiretap channel (WTC$_2$ for short) denoted by $\left(\mathcal{X}_1, \mathcal{X}_2, \mathcal{Y}_1, \mathcal{Y}_2, \mathcal{Z}, p_{Y_1 Y_2 Z | X_1 X_2}\right)$ is a two-way wiretap channel for which:

$$\forall (x_1^n, x_2^n, y_1^n, y_2^n, z^n) \in \mathcal{X}_1^n \times \mathcal{X}_2^n \times \mathcal{Y}_1^n \times \mathcal{Y}_2^n \times \mathcal{Z}^n,$$
$$p_{Y_1^n Y_2^n Z^n | X_1^n X_2^n}(y_1^n, y_2^n, z^n | x_1^n, x_2^n) =$$
$$\prod_{i=1}^n p_{Y_1 Y_2 Z | X_1 X_2}\left(y_1^{(i)}, y_2^{(i)}, z^{(i)} | x_1^{(i)}, x_2^{(i)}\right). \quad (2)$$

Now, we formally define a code for the two-way wiretap channel.

*Definition 3:* A $\left(2^{nR_1}, 2^{nR_2}, n\right)$ two-way wiretap channel code $\mathcal{C}_n$ consists of:

1) Two message alphabets: $\mathcal{M}_1 = [\![1, 2^{nR_1}]\!]$ and $\mathcal{M}_2 = [\![1, 2^{nR_2}]\!]$.
2) Two local sources of randomness $(\mathcal{R}_1, p_{R_1})$ and $(\mathcal{R}_2, p_{R_2})$ independent of the channel and messages.
3) Two sets of encoding functions that map a message and past channel observations to a channel input symbol:
   - $n$ encoding functions for transmitter 1:
   $$\forall i \in [\![1, n]\!], \, f_1^{(i)} : \mathcal{M}_1 \times \mathcal{Y}_1^{i-1} \times \mathcal{R}_1 \to \mathcal{X}_1;$$
   - $n$ encoding functions for transmitter 2:
   $$\forall i \in [\![1, n]\!], \, f_2^{(i)} : \mathcal{M}_2 \times \mathcal{Y}_2^{i-1} \times \mathcal{R}_2 \to \mathcal{X}_2.$$
4) Two decoding functions that map channel observations to a message or an error symbol "?":
   - $g_1 : \mathcal{Y}_1^n \times \mathcal{M}_1 \times \mathcal{R}_1 \to \mathcal{M}_2 \cup \{?\}$;
   - $g_2 : \mathcal{Y}_2^n \times \mathcal{M}_2 \times \mathcal{R}_2 \to \mathcal{M}_1 \cup \{?\}$.

We also define the following quantities to assess the performance of a code $\mathcal{C}_n$:

- the probability of error:
$$P_e(\mathcal{C}_n) \triangleq \mathbb{P}\left((M_1, M_2) \neq (\hat{M}_1, \hat{M}_2) | \mathcal{C}_n\right);$$

- the information leakage to the eavesdropper:
$$L(\mathcal{C}_n) \triangleq \mathbb{I}(Z^n; M_1 M_2 | \mathcal{C}_n).$$

*Definition 4:* A rate pair $(R_1, R_2)$ is achievable for a WTC$_2$ if there exists a sequence of codes $\{\mathcal{C}_n\}_{n \geqslant 1}$ such that:

- $\lim_{n \to \infty} P_e(\mathcal{C}_n) = 0$ (Reliability);
- $\lim_{n \to \infty} L(\mathcal{C}_n) = 0$ (Strong secrecy).

*Definition 5:* The *strong secrecy capacity* region $\bar{\mathcal{R}}^{2\text{W}}$ is defined as:

$$\bar{\mathcal{R}}^{2\text{W}} \triangleq \text{cl}\left(\{(R_1, R_2) : (R_1, R_2) \text{ is achievable}\}\right).$$

We denote by $\mathcal{R}^{2\text{W}}$ the *weak* secrecy capacity region.

We are interested in characterizing the entire region of achievable rate pairs $(R_1, R_2)$, but it is rather difficult to obtain a closed-form expression. In principle, the coding scheme in Definition 3 could simultaneously exploit the *interference* of transmitted signals at the eavesdropper's terminal and *feedback*. To obtain some insight, we study instead a simpler strategy that partially *decouples* these two effects as follows:

- First, we exploit interference to penalize the eavesdropper and increase secure communication rates. The interference can be of two types: interference between codewords or jamming with noise.
- Next, we introduce feedback by means of *key exchange* and *key generation*. With key exchange, one user sacrifices part of its secure communication rate to *exchange* a secret key, whereas with key generation both users exploit channel randomness to *distill* keys. Those keys are then used to encrypt messages with a one-time pad.

## III. RESOLVABILITY-BASED COOPERATIVE JAMMING

### A. Cooperative Jamming

A natural attempt to increase secure communication rates consists in jamming Eve with noise, in order to decrease her signal-to-noise ratio. This strategy, called *cooperative jamming* [23], forces one user to stop transmitting information to jam the eavesdropper. To overcome this limitation, Alice and Bob can use codewords whose interference also has a detrimental effect on Eve without sacrificing as much information rate. This scheme is called *coded cooperative jamming* and was introduced by Tekin and Yener [7], [8]. It is possible to combine both strategies and have Alice and Bob perform coded cooperative jamming while simultaneously jamming the eavesdropper with noise. Simultaneous cooperative jamming can be implemented by prefixing an artificial discrete memoryless channel (DMC) before the $\text{WTC}_2$ and sending codewords through the concatenated channels. This technique is therefore called *prefixing* in [11].

Note that cooperative jamming does not exploit feedback, which corresponds to using only two encoding functions $f_1^{(1)}$ and $f_2^{(1)}$ in Definition 3. Using cooperative jamming is then equivalent to studying the simplified channel model illustrated in Fig. 2, in which the eavesdropper observes the output of a multiple-access channel.

### B. Achievable Region

We first derive an achievable region using the notion of channel resolvability and *auxiliary* messages $M_1' \in \mathcal{M}_1' \triangleq [\![1, 2^{nR_1'}]\!]$, uniformly distributed, and $M_2' \in \mathcal{M}_2' \triangleq [\![1, 2^{nR_2'}]\!]$, uniformly distributed, in place of $(\mathcal{R}_1, p_{R_1})$ and $(\mathcal{R}_2, p_{R_2})$. This region is given in Proposition 1.

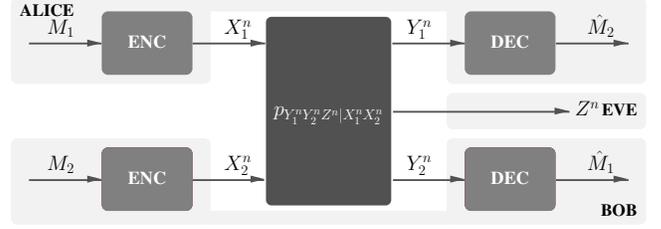

Fig. 2. Communication over a two-way wiretap channel without feedback

*Proposition 1:*

$$\mathcal{R} = \underset{R_1, R_2}{\text{Proj}} \bigcup_{p \in \mathcal{P}} \left\{ \begin{pmatrix} R_1 \\ R_2 \\ R_1' \\ R_2' \end{pmatrix} \middle| \begin{array}{c} R_1 + R_1' \leqslant \mathbb{I}(Y_2; C_1 | X_2) \\ R_2 + R_2' \leqslant \mathbb{I}(Y_1; C_2 | X_1) \\ R_1' + R_2' \geqslant \mathbb{I}(C_1 C_2; Z) \\ R_1' \geqslant \mathbb{I}(C_1; Z) \\ R_2' \geqslant \mathbb{I}(C_2; Z) \end{array} \right\} \subset \bar{\mathcal{R}}^{2\text{W}}, \quad (3)$$

where $\text{Proj}_{R_1, R_2}$ is the projection on the plane of rates $(R_1, R_2)$, and:

$$\mathcal{P} = \{p_{X_1 X_2 C_1 C_2 Y_1 Y_2 Z} \text{ factorizing as: } p_{Y_1 Y_2 Z | X_1 X_2} \, p_{X_1 | C_1} \, p_{C_1} \, p_{X_2 | C_2} \, p_{C_2}\}. \quad (4)$$

*Remark* A similar result has been independently established by Yassaee and Aref [24] using a related technique based on approximation of output statistics. However, their proof only holds for discrete memoryless channel because it is based on strongly typical sequences. In our proof, we use Steinberg's results [19], which also hold for Gaussian memoryless channels.

*Proof of Proposition 1:*

In this proof, we consider two types of transmitted messages. First, the main messages, which must be transmitted between Alice and Bob reliably and securely with respect to Eve. Second, the auxiliary messages, which are used to perform coded cooperative jamming and introduce randomness to mislead the eavesdropper. Although auxiliary messages do not carry information by themselves, we require that Alice and Bob decode them *reliably*. We also perform simultaneous cooperative jamming by prefixing DMCs.

The proof is based on random coding. We fix distributions $p_{C_1}$, $p_{C_2}$, $p_{X_1 | C_1}$ and $p_{X_2 | C_2}$. Let $\epsilon > 0$.

*Code creation:* We create the code as follows:

- **Code generation:** we generate the codewords at random: Generate $\lceil 2^{nR_1} \rceil \lceil 2^{nR_1'} \rceil$ i.i.d. sequences $c_1^n(\mu_1)$ with $\mu_1 = (i, j) \in \mathcal{M}_1 \times \mathcal{M}_1'$ according to $p_{C_1}$, and $\lceil 2^{nR_2} \rceil \lceil 2^{nR_2'} \rceil$ i.i.d. sequences $c_2^n(\mu_2)$ with $\mu_2 = (i, j) \in \mathcal{M}_2 \times \mathcal{M}_2'$ according to $p_{C_2}$. Here, $i$ represents the index of the main message, $j$ the index of the auxiliary message. We denote by $\mathcal{C}_n$ the random variable representing the generated code and by $\mathcal{C}_n$ one of its realizations.
- **Encoding:** If Alice wants to send $\mu_1 \in \mathcal{M}_1 \times \mathcal{M}_1'$, she computes $c_1^n(\mu_1)$, and transmits $x_1^n$ obtained by sending $c_1^n$ through a DMC with transition probability $p_{X_1 | C_1}$.
  Similarly, if Bob wants to send $\mu_2 \in \mathcal{M}_2 \times \mathcal{M}_2'$, he computes $c_2^n(\mu_2)$, and transmits $x_2^n$ obtained by sending $c_2^n$ through a DMC with transition probability $p_{X_2 | C_2}$.

The DMCs with transition probabilities $p_{X_1|C_1}$ and $p_{X_2|C_2}$ are prefix channels used for simultaneous cooperative jamming to confuse the eavesdropper.

- **Decoding:** we consider a typical set decoder.
  Let us define $\mathcal{A}_{1,\epsilon}^n \triangleq \mathcal{A}_\epsilon^n(X_1, C_2, Y_1)$ and $\mathcal{A}_{2,\epsilon}^n \triangleq \mathcal{A}_\epsilon^n(X_2, C_1, Y_2)$.
  If $y_1^n$ is received by Alice, she selects $\hat{\mu}_2$ such that $(x_1^n, c_2^n(\hat{\mu}_2), y_1^n) \in \mathcal{A}_{1,\epsilon}^n$. If such a tuple exists and is unique, output $\hat{\mu}_2$, otherwise declare an error ($\hat{\mu}_2 = ?$).
  If $y_2^n$ is received by Bob, he selects $\hat{\mu}_1$ such that $(x_2^n, c_1^n(\hat{\mu}_1), y_2^n) \in \mathcal{A}_{2,\epsilon}^n$. If such a tuple exists and is unique, output $\hat{\mu}_1$, otherwise declare an error ($\hat{\mu}_1 = ?$).

*Probability of error analysis:* We study the probability of error defined as:

$$P_e(\mathcal{C}_n) \triangleq \mathbb{P}\Big((M_1, M_2, M_1', M_2') \neq (\hat{M}_1, \hat{M}_2, \hat{M}_1', \hat{M}_2')\Big|\mathcal{C}_n\Big). \quad (5)$$

We are looking for conditions on rates such that this probability of error goes to zero as $n$ goes to infinity. We use the following Lemma.

*Lemma 1:* (Probability of error) For $\epsilon > 0$,

$$\begin{cases} R_1 + R_1' < \mathbb{I}(Y_2; C_1|X_2) \\ R_2 + R_2' < \mathbb{I}(Y_1; C_2|X_1) \end{cases} \Rightarrow \lim_{n \to \infty} \mathbb{E}(P_e(C_n)) \leqslant \delta(\epsilon). \quad (6)$$

*Proof:* see Appendix A. ∎

*Remark* Lemma 1 gives conditions for reliable communication regardless of the eavesdropper, which is intuitive because reliability only depends on what is happening between the two legitimate users. However, this differs from the proof in [7], in which additional reliability constraints that depend on the eavesdropper are introduced to compute the leakage.

*Leakage analysis:* We study the leakage defined as:

$$L(\mathcal{C}_n) \triangleq \mathbb{I}(Z^n; M_1 M_2 | \mathcal{C}_n). \quad (7)$$

*Lemma 2:* (Leakage) For $\epsilon > 0$,

$$\begin{cases} R_1' + R_2' > \mathbb{I}(C_1 C_2; Z) \\ R_1' > \mathbb{I}(C_1; Z) \\ R_2' > \mathbb{I}(C_2; Z) \end{cases} \Rightarrow \lim_{n \to \infty} \mathbb{E}(L(C_n)) \leqslant \delta(\epsilon). \quad (8)$$

*Proof:* see Appendix B. ∎

*Remark* To interpret this result, we need to precisely understand the role of auxiliary messages $M_1'$ and $M_2'$. These messages are introduced to replace the sources of randomness and Lemma 2 gives lower bounds on the rates of these auxiliary messages. This result is intuitive: to prevent Eve from recovering the messages, more randomness must be introduced in the encoding process. Lemma 2 shows that there exist minimum values of the rates such that they allow zero asymptotical leakage.

It is important to remember that, because of the constraints imposed by Lemma 1, increasing auxiliary message rates reduces the amount of information one can transmit through the channel.

*Code selection:* By Markov's inequality, we have

$$\lim_{n \to \infty} \mathbb{P}(P_e(C_n) > 3\delta(\epsilon)) \leqslant \frac{1}{3},$$
$$\lim_{n \to \infty} \mathbb{P}(L(C_n) > 3\delta(\epsilon)) \leqslant \frac{1}{3};$$

therefore, we conclude that there exists a specific sequence of codes $\{C_n\}_{n \geqslant 1}$ such that $\lim_{n \to \infty} P_e(C_n) \leqslant \delta(\epsilon)$, and $\lim_{n \to \infty} L(C_n) \leqslant \delta(\epsilon)$.

*Remark* Each code $\mathcal{C}_n$ consists of a pair of codes $(\mathcal{C}_1, \mathcal{C}_2)$. Although the codes $\mathcal{C}_1$ and $\mathcal{C}_2$ are generated according to independent distributions, note that both codes are *jointly* selected; therefore, the codes are used independently by Alice and Bob but optimized jointly to guarantee secrecy.

*Conclusion:* For any $\epsilon > 0$, there exists a code such that the probability of error and the leakage are smaller than $\delta(\epsilon)$. Therefore, it is possible to create a sequence of codes $\{C_n(\epsilon_n)\}_{n \geqslant 1}$ with $\epsilon_n \xrightarrow[n \to \infty]{} 0$.

By combining the rate constraints in (6) and (8), we get the result given in Proposition 1. ∎

By elimination of the auxiliary rates $R_1'$ and $R_2'$, we obtain the following region:

*Corollary 1 (Strongly secure achievable region):*

$$\mathcal{R} = \bigcup_{p \in \mathcal{P}} \left\{ \begin{pmatrix} R_1 \\ R_2 \end{pmatrix} \middle| \begin{array}{l} R_1 \leqslant \mathbb{I}(Y_2; C_1|X_2) - \mathbb{I}(C_1; Z) \\ R_2 \leqslant \mathbb{I}(Y_1; C_2|X_1) - \mathbb{I}(C_2; Z) \\ R_1 + R_2 \leqslant \mathbb{I}(Y_2; C_1|X_2) + \mathbb{I}(Y_1; C_2|X_1) \\ \qquad\qquad\qquad - \mathbb{I}(C_2 C_2; Z) \end{array} \right\}$$
$$\subset \bar{\mathcal{R}}^{2W}, \quad (9)$$

where:

$$\mathcal{P} = \{p_{X_1 X_2 C_1 C_2 Y_1 Y_2 Z} \text{ factorizing as:}$$
$$p_{Y_1 Y_2 Z | X_1 X_2} p_{X_1|C_1} p_{C_1} p_{X_2|C_2} p_{C_2}\}. \quad (10)$$

*Proof of Corollary 1:* With (3) and (8), we have:

$$R_1 + R_1' + R_2 + R_2' \leqslant \mathbb{I}(Y_2; C_1|X_2) + \mathbb{I}(Y_1; C_2|X_1)$$
$$R_1 + R_2 \leqslant \mathbb{I}(Y_2; C_1|X_2) + \mathbb{I}(Y_1; C_2|X_1) - R_1' - R_2'$$
$$R_1 + R_2 \leqslant \mathbb{I}(Y_2; C_1|X_2) + \mathbb{I}(Y_1; C_2|X_1) - \mathbb{I}(C_2 C_2; Z).$$

Using (3) and (8), we also have:

$$R_1 + R_1' \leqslant \mathbb{I}(Y_2; C_1|X_2)$$
$$R_1 \leqslant \mathbb{I}(Y_2; C_1|X_2) - R_1'$$
$$R_1 \leqslant \mathbb{I}(Y_2; C_1|X_2) - \mathbb{I}(C_1; Z).$$

Similarly, we have: $R_2 \leqslant \mathbb{I}(Y_1; C_2|X_1) - \mathbb{I}(C_2; Z)$. ∎

The region described in (9) is identical to the one obtained by Tekin and Yener in [7], [8] for weak secrecy. However, a closer look at the proof shows that their result is obtained by projecting the region $\mathcal{R}'$ defined as:

$$\mathcal{R}' = \bigcup_{p \in \mathcal{P}} \left\{ \begin{pmatrix} R_1 \\ R_2 \\ R'_1 \\ R'_2 \end{pmatrix} \middle| \begin{array}{c} R_1 + R'_1 \leqslant \mathbb{I}(Y_2; C_1 | X_2) \\ R_2 + R'_2 \leqslant \mathbb{I}(Y_1; C_2 | X_1) \\ R'_1 + R'_2 = \mathbb{I}(C_1 C_2; Z) \\ R'_1 \leqslant \mathbb{I}(C_1; Z | C_2) \\ R'_2 \leqslant \mathbb{I}(C_2; Z | C_1) \end{array} \right\}. \quad (11)$$

This region differs from the one in Proposition 1 in the constraints on auxiliary message rates $(R'_1, R'_2)$. The difference is shown in Fig. 3, where the dark area corresponds to constraints (8) and the light one to the constraints on $(R'_1, R'_2)$ in (11). Note that the latter corresponds to the achievable region of a multiple access channel (MAC). This is not surprising because the proof of (11) relies explicitly on the analysis of the probability of error for the eavesdropper, who obtains its signal through a virtual multiple access channel as shown in Fig. 2. In contrast, our approach analyzes the secrecy constraint directly, which leads to lower bounds on the auxiliary message rate required to confuse the eavesdropper. The projection of both regions on the plane of rates $(R_1, R_2)$ is the same because it corresponds to having the auxiliary message rates on the diagonal edge, which is common to both. In terms of code structure, the approach of Tekin and Yener consists in augmenting the number of auxiliary messages until the leakage to the eavesdropper become negligible, which only happens on the slope of the MAC region. Our constraints directly give a region with negligible leakage and the approach consists in finding the minimum number of auxiliary messages needed to confuse the eavesdropper, and, therefore, augmenting the number of secret messages to the maximum possible value.

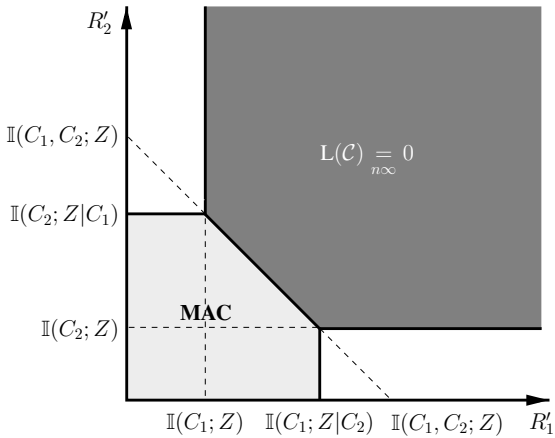

Fig. 3. Constraints on $R'_1$ and $R'_2$ in (8) and (11)

## IV. KEY EXCHANGE AND KEY GENERATION

In the previous section, we have exploited the benefits of coded cooperative jamming and simultaneous cooperative jamming, but we did not exploit the possibility of feedback. We introduce feedback with two mechanisms:

- we use the technique of [9] and [11] to transfer secret rate from one user to the other;
- we use the randomness introduced by cooperative jamming to induce a source from which we generate secret keys.

Note that we can only exchange information through a rate-limited channel; therefore, we rely on results for secret key agreement with rate-limited public communication [16], [17].

### A. Key Exchange

We introduce key exchange on top of the cooperative jamming scheme introduced in Section III by splitting the main and auxiliary messages into multiple parts. We use these sub-messages as a means to exchange a key using the secret channel and encrypt part of the public message. Because of the "secret rate transfer" we need to redefine which rates we are considering.

Let us consider a code for cooperative jamming with secret message rates $(R_1, R_2)$ and auxiliary messages rates $(R'_1, R'_2)$. For $i \in \{1, 2\}$, we split the main message $M_i$ into two parts:

- A *key*, which is used for encryption by the other user: $K_i \in \mathcal{K}_i = [\![1, 2^{nR_i^k}]\!]$; user $i$ needs to sacrifice a part of its secret message to transmit this key.
- A *secret* message: $S_i \in \mathcal{S}_i = [\![1, 2^{nR_i^s}]\!]$; this part corresponds to the part of secret message user $i$ does not sacrifice.

We also split the auxiliary message $M'_i$ in two parts:

- A message $E_i$, which is *encrypted* with the key sent by the other user to give $\bar{E}_i \in \mathcal{E}_i = [\![1, 2^{nR_i^e}]\!]$. Encryption is done with a one-time pad to ensure perfect secrecy [25].
- An *open* message: $O_i \in \mathcal{O}_i = [\![1, 2^{nR_i^o}]\!]$; this corresponds to the part of the message that remains public and still perfectly decodable by the receiver.

By convention, if no key is available $R_i^e = 0$, else, by construction, $R_1^e \leqslant R_2^k$, $R_2^e \leqslant R_1^k$. We denote: $\mathcal{M}_i = \mathcal{S}_i \times \mathcal{K}_i$, $\mathcal{M}'_i = \mathcal{O}_i \times \mathcal{E}_i$. Therefore, we have the following relationships between the various rates:

$$R_1 = R_1^s + R_1^k, \quad R_2 = R_2^s + R_2^k,$$
$$R'_1 = R_1^o + R_1^e, \quad R'_2 = R_2^o + R_2^e. \quad (12)$$

Although $(R_1, R_2, R'_1, R'_2)$ still represents the rates provided by cooperative jamming, they are no longer the rates of interest after key exchange. In fact, part of the auxiliary message is encrypted while part of the secret message is sacrificed to exchange a key. Hence, the rates we need to consider are the following:

- a pair of secret rates: $\tilde{R}_1 = R_1^s + R_1^e$, and $\tilde{R}_2 = R_2^s + R_2^e$;
- a pair of public rates: $\tilde{R}'_1 = R_1^o$, and $\tilde{R}'_2 = R_2^o$.

*Remark* Because the key sent by one user cannot be used simultaneously by the other, the key exchange scheme must operate in several rounds. At each round, we use the key sent in the previous one, except in the first round where no key is available. If we use a code of length $n$ $B$ times, we obtain a new code of length $n' = Bn$; the first message does not use key exchange, but the next $B - 1$ do. If the communication rate for the first message is $R^*$, for the $B - 1$ others it is $R$, and the overall rate is $\bar{R}$, we have:

$$\bar{R} = \frac{nR^* + n(B-1)R}{nB} \underset{B \to \infty}{=} R.$$

Thus, the first round incurs a negligible rate penalty for $B$ going to infinity.

For weak secrecy, [11] proves the following proposition:

*Proposition 2 (El Gamal et al.):*

$$\mathcal{R}^{\mathrm{F}} = \bigcup_{p \in \mathcal{P}} \left\{ \begin{pmatrix} R_1 \\ R_2 \\ R'_1 \\ R'_2 \end{pmatrix} \middle| \begin{array}{c} R_1 \leqslant \mathbb{I}(Y_2; C_1 | X_2) \\ R_2 \leqslant \mathbb{I}(Y_1; C_2 | X_1) \\ R_1 + R_2 \leqslant \mathbb{I}(Y_2; C_1 | X_2) + \mathbb{I}(Y_1; C_2 | X_1) \\ - \mathbb{I}(C_2 C_2; Z) \end{array} \right\} \subseteq \mathcal{R}^{2\mathrm{W}}, \quad (13)$$

where:

$$\mathcal{P} = \{ p_{X_1 X_2 C_1 C_2 Y_1 Y_2 Z} \text{ factorizing as:} \\ p_{Y_1 Y_2 Z X_1 X_2} \, p_{X_1 | C_1} \, p_{C_1} \, p_{X_2 | C_2} \, p_{C_2} \}. \quad (14)$$

Based on Section III, we also have this result for strong secrecy: $\mathcal{R}^{\mathrm{F}} \subset \bar{\mathcal{R}}^{2\mathrm{W}}$.

*Remark* By comparing (9) and (13), we see that key exchange improves the individual bounds on $R_1$ and $R_2$, but not the bound on the sum-rate. Individual bounds on $R_1$ and $R_2$ actually correspond to the capacity of the channel between the two users and cannot be improved; therefore, any improvement on the region should modify the sum-rate constraint.

### B. Key Generation from Induced Source

Key exchange requires sacrificing part of the secret rate of one user. In addition, we can use the channel randomness introduced for simultaneous cooperative jamming to *extract* secret keys. Although users must exchange additional messages to agree on a common secret key, key generation only comes at the expense of *public* message rate.

In the next section, we study the Gaussian two-way wiretap channel and we explicitly show how cooperative jamming induces a discrete memoryless source (DMS) $(\tilde{X}, \tilde{Y}, \tilde{Z})$ independent of all other observations, which can be used to distill a secret key. The existence of this DMS is not obvious because the noise introduced by cooperative jamming is already exploited to harm the eavesdropper and it is *a priori* unclear if it could be used simultaneously as a source of common randomness for secret-key agreement. We start by clarifying what knowledge of the eavesdropper needs to be considered to identify the DMS. Let $K$ be the secret key to generate and let $M_1$, $M_2$, $M'_1$ and $M'_2$ be the secret and auxiliary messages used in the cooperative jamming code. The key and secret messages must be independent and both must be hidden from the eavesdropper, that is for $\epsilon > 0$, $\mathbb{I}(M_1 M_2 K; Z^n) \leqslant \epsilon$. Notice that

$$\mathbb{I}(M_1 M_2 K; Z^n) = \mathbb{I}(M_1 M_2; Z^n) + \mathbb{I}(K; Z^n | M_1 M_2) \\ = \mathbb{I}(M_1 M_2; Z^n) + \mathbb{I}(K; Z^n M_1 M_2).$$

The term $\mathbb{I}(M_1 M_2; Z^n)$ can be smaller than $\epsilon/2$ by construction of the cooperative jamming code but note that we must then have $\mathbb{I}(K; Z^n M_1 M_2) \leqslant \epsilon/2$; this means that the key must be hidden from an eavesdropper having access not only to the observation $Z^n$ but also to the secret messages $M_1$ and $M_2$. Furthermore, note that

$$\mathbb{I}(M_1 M_2 K; Z^n) \\ = \mathbb{I}(M_1 M_2; Z^n) + \mathbb{I}(K; Z^n M_1 M_2 M'_1 M'_2) \\ - \mathbb{I}(K; M'_1 M'_2 | Z^n M_1 M_2).$$

If the cooperative jamming code is chosen to provide the highest secrecy rate, the public message rates must be chosen to lie on the boundary of the region in (9), for which [7], [8] shows that $\mathbb{H}(M'_1 M'_2 | Z^n M_1 M_2) \leqslant \epsilon/2$. In this case, we obtain:

$$\mathbb{I}(M_1 M_2 K; Z^n) \geqslant \mathbb{I}(M_1 M_2; Z^n) \\ + \mathbb{I}(K; Z^n M_1 M_2 M'_1 M'_2) - \frac{\epsilon}{2}$$

In other words, we can assume that the key $K$ must be kept secret from an eavesdropper observing not only the channel output $Z^n$ but also knowing the secret and auxiliary messages transmitted by the legitimate users.

### C. Achievable region with key exchange and key generation

In this section, we provide a generic strategy and an achievable region based on cooperative jamming, key exchange and key generation. We assume the existence of a DMS $(\tilde{X}, \tilde{Y}, \tilde{Z})$ independent of all other observations induced by cooperative jamming. The statistics of the DMS depend both on channel statistics and the code used for communications, but we do not attempt to characterize them exactly.

First, note that substituting the rates defined in (12) in Proposition 1 gives a set of constraints that achievable rates must satisfy:

$$R_1^{\mathrm{s}} + R_1^{\mathrm{k}} + R_1^{\mathrm{o}} + R_1^{\mathrm{e}} \leqslant \mathbb{I}(Y_2; C_1 | X_2) \quad (15)$$
$$R_2^{\mathrm{s}} + R_2^{\mathrm{k}} + R_2^{\mathrm{o}} + R_2^{\mathrm{e}} \leqslant \mathbb{I}(Y_1; C_2 | X_1) \quad (16)$$
$$R_1^{\mathrm{o}} + R_1^{\mathrm{e}} + R_2^{\mathrm{o}} + R_2^{\mathrm{e}} \geqslant \mathbb{I}(C_1 C_2; Z) \quad (17)$$
$$R_1^{\mathrm{o}} + R_1^{\mathrm{e}} \geqslant \mathbb{I}(C_1; Z) \quad (18)$$
$$R_2^{\mathrm{o}} + R_2^{\mathrm{e}} \geqslant \mathbb{I}(C_2; Z). \quad (19)$$

In order to add a key generation step, we need to further split message rates in multiple parts. For instance, Alice modifies her rates as follows:

- $\tilde{R}_1^{\mathrm{s}} = R_1^{\mathrm{s}}$. Secret key generation does not change anything for the secret message rate.
- $\tilde{R}_1^{\mathrm{k}} = R_1^{\mathrm{k}}$. Secret key generation does not change anything for the secret key rate used for key exchange.
- $\tilde{R}_1^{\mathrm{o}} = R_1^{\mathrm{o}} - \bar{R}_1^{\mathrm{p}} - \bar{R}_1^{\mathrm{e}}$. A part $\bar{R}_1^{\mathrm{p}}$ of the open message rate is used for secret-key generation while a part $\bar{R}_1^{\mathrm{e}}$ is used to transmit an encrypted message with that same key. Notice that the constraint $R_1^{\mathrm{o}} \geqslant \bar{R}_1^{\mathrm{p}} + \bar{R}_1^{\mathrm{e}}$ must be satisfied.
- $\tilde{R}_1^{\mathrm{e}} = R_1^{\mathrm{e}} + \bar{R}_1^{\mathrm{e}}$. Thanks to key generation the encrypted message rate increases.

Alice's total secure communication rate is then $\tilde{R}_1^{\mathrm{s}} + \tilde{R}_1^{\mathrm{e}}$. Bob modifies his rates in a similar way.

Even if the DMS was characterized exactly, the secret-key capacity of a source with rate-limited public communication is not known. Therefore, we focus on a suboptimal key agreement strategy in which a single user sacrifices a fraction $\bar{R}^{\mathrm{p}}$ of its open message rate for communication and a single user uses the secret-key generated from the source for encryption. In this case, the optimal key generation rate $\bar{R}^k$ that can be distilled with rate-limited public communication is known [16, Theorem 2.6]:

$$\bar{R}^k < \mathbb{I}\left(V; \tilde{X}|U\right) - \mathbb{I}\left(V; \tilde{Z}|U\right),$$

where the random variables $U$ and $V$ are such that $U \leftrightarrow V \leftrightarrow \tilde{Y} \leftrightarrow \tilde{X}\tilde{Z}$ forms a Markov chain and

$$\mathbb{I}\left(V; \tilde{Y}\right) - \mathbb{I}\left(V; \tilde{X}\right) \leqslant \bar{R}^p.$$

Depending on which user performs public discussion and which user encrypts a message we obtain additional constraints on $\bar{R}^p$ and $\bar{R}^k$. Specifically, we have four possible configurations given in Table I.

| encryption \ public com. | Alice | Bob |
|---|---|---|
| Alice: $\bar{R}_1^e = \bar{R}_1^k$, $\bar{R}_2^k = 0$ | $\bar{R}_1^p + \bar{R}_1^k \leqslant R_1^o$ | $\begin{cases} \bar{R}_1^p \leqslant R_2^o \\ \bar{R}_1^k \leqslant R_1^o \end{cases}$ |
| Bob: $\bar{R}_2^e = \bar{R}_2^k$, $\bar{R}_1^k = 0$ | $\begin{cases} \bar{R}_2^p \leqslant R_1^o \\ \bar{R}_2^k \leqslant R_2^o \end{cases}$ | $\bar{R}_2^p + \bar{R}_2^k \leqslant R_2^o$ |

TABLE I
ADDITIONAL CONSTRAINTS FOR KEY GENERATION

Although we do not attempt to obtain a closed-form expression, note that key generation can loosen the sum-rate constraint in (13).

## V. GAUSSIAN TWO-WAY WIRETAP CHANNEL

In this section, we evaluate an achievable region based on the strategies developed in previous sections. We consider a Gaussian two-way wiretap channel for which the relationships between inputs and outputs are:

$$\begin{cases} Y_1^n = \sqrt{g_1}X_1^n + X_2^n + N_{21}^n \\ Y_2^n = X_1^n + \sqrt{g_2}X_2^n + N_{12}^n \\ Z^n = \sqrt{h_1}X_1^n + \sqrt{h_2}X_2^n + N_e^n. \end{cases} \quad (20)$$

In addition the inputs are subject to the power constraints:

$$\frac{1}{n}\sum_{i=1}^{n}\mathbb{E}\left(\left(X_1^{(i)}\right)^2\right) \leqslant \rho_1 \text{ and } \frac{1}{n}\sum_{i=1}^{n}\mathbb{E}\left(\left(X_2^{(i)}\right)^2\right) \leqslant \rho_2.$$

We suppose the additive noises $N_{21}^n$, $N_{12}^n$, and $N_e^n$ are i.i.d. zero-mean unit-variance Gaussian vectors. We implement cooperative jamming by prefixing an additive white Gaussian noise (AWGN) channel before each input of $WTC_2$. This is equivalent to considering a modified channel with inputs $C_1$ and $C_2$, with:

$$X_1^n = C_1^n + N_{11}^n \quad \text{and} \quad X_2^n = C_2^n + N_{22}^n, \quad (21)$$

where $N_{11}^n \sim \mathcal{N}(0, \rho_1^n \mathbf{I}_n)$ and $N_{22}^n \sim \mathcal{N}(0, \rho_2^n \mathbf{I}_n)$ correspond to the noise introduced by users to realize cooperative jamming. We suppose $\rho_1^n \leqslant \rho_1$ and $\rho_2^n \leqslant \rho_2$, and obtain the modified power constraints:

$$\frac{1}{n}\sum_{i=1}^{n}\mathbb{E}\left(\left(C_k^{(i)}\right)^2\right) \leqslant \rho_k - \rho_k^n = \rho_k^c, \ k = 1, 2.$$

With minor modifications of the proof of Section III to account for the power constraints, we obtain an achievable region for the Gaussian two-way wiretap channel by substituting the random variables $C_1 \sim \mathcal{N}(0, \rho_1^c)$ and $C_2 \sim \mathcal{N}(0, \rho_2^c)$ in the bounds obtained earlier.

### A. Randomness Source Extraction

We provide an explicit characterization of the DMS induced by cooperative jamming, which we will use to extract a key later on. Define:

$$\begin{aligned}\tilde{X}_1 &= N_{11} & \tilde{Y}_1 &= \tilde{X}_2 + N_{21} \\ \tilde{X}_2 &= N_{22} & \tilde{Y}_2 &= \tilde{X}_1 + N_{12}\end{aligned} \quad \tilde{Z} = \sqrt{h_1}\,\tilde{X}_1 + \sqrt{h_2}\,\tilde{X}_2 + N_e. \quad (22)$$

*Proposition 3:* The triple $(\tilde{X}, \tilde{Y}, \tilde{Z})$ with $\tilde{X} = (\tilde{X}_1, \tilde{Y}_1)$ and $\tilde{Y} = (\tilde{X}_2, \tilde{Y}_2)$ is an independent DMS that can be used for key generation. Alice, Bob and Eve observe the components $\tilde{X}$, $\tilde{Y}$, and $\tilde{Z}$, respectively.

*Proof:* Since Alice generates the noise $N_{11}$, she obtains $\tilde{X}_1$ directly. She also observes the channel output $Y_1^n = C_2^n + N_{22}^n + N_{21}^n + \sqrt{g_1}\,X_1^n$; since she knows $X_1^n$ and can decode $C_2^n$ with high probability, she can obtain $\tilde{Y}_1$. Similarly, Bob can obtain $\tilde{X}_2$ and $\tilde{Y}_2$.

Eve observes the channel output $Z^n = \sqrt{h_1}\,X_1^n + \sqrt{h_2}\,X_2^n + N_e^n$ and, as discussed in Section IV, we assume she obtains $C_1^n$ and $C_2^n$ as side-information. Therefore, Eve can compute $\tilde{Z}$, but we also need to show that $\tilde{Z}^n$ is a sufficient statistic for $\tilde{X}^n$ and $\tilde{Y}^n$ given $(Z^n, C_1^n, C_2^n)$. This is the case because:

$$\mathbb{H}(\tilde{X}^n, \tilde{Y}^n | Z^n C_1^n C_2^n) \stackrel{(a)}{=} \mathbb{H}(\tilde{X}^n, \tilde{Y}^n | \tilde{Z}^n C_1^n C_2^n)$$
$$\stackrel{(b)}{=} \mathbb{H}(\tilde{X}^n, \tilde{Y}^n | \tilde{Z}^n)$$

because (a): $ZC_1C_2 \mapsto \tilde{Z}C_1C_2$ is bijective, and (b): $(\tilde{X}, \tilde{Y}, \tilde{Z})$ is independent of $(C_1, C_2)$. ∎

The DMS $(\tilde{X}, \tilde{Y}, \tilde{Z})$ is a vector source, and we do not have a closed-form expression for the secret key capacity with rate limited public-rate communication. However, if Alice and Bob ignore one of their observations, the DMS reduces to a degraded scalar source. By [17, Corollary 2], there exists a function $f$ with closed-form expression such that:

- If Alice ignores $\tilde{Y}_1$ and Bob ignores $\tilde{X}_2$, we have $\tilde{Y}_2 \leftrightarrow \tilde{X}_1 \leftrightarrow \tilde{Z}$. In this case, if Bob sends public messages at rate $\bar{R}_p$, Alice and Bob can distill a key at rate $\bar{R}_k$, such that $\bar{R}_k \leqslant f\left(\bar{R}_p, p_{\tilde{Y}_2\tilde{X}_1\tilde{Z}}\right)$.
- If Bob ignores $\tilde{Y}_2$ and Alice ignores $\tilde{X}_1$, we have $\tilde{Y}_1 \leftrightarrow \tilde{X}_2 \leftrightarrow \tilde{Z}$. In this case, if Alice sends public messages at rate $\bar{R}_p$, Alice and Bob can distill a key at rate $\bar{R}_k$, such that $\bar{R}_k \leqslant f\left(\bar{R}_p, p_{\tilde{Y}_1\tilde{X}_2\tilde{Z}}\right)$.

With the explicit characterization of the DMS induced by cooperative jamming and the function $f$, it is possible to compute new achievable rates for the strategy described in Section IV. Although we do not have a closed-form expression for the resulting region $\bar{\mathcal{R}}^K$, it is possible to prove that the region strictly includes the region $\bar{\mathcal{R}}^F$ in Proposition 2 strengthened for strong secrecy.

*Proposition 4:* Let us consider $\bar{\mathcal{R}}^F$, $\bar{\mathcal{R}}^K$, and $\bar{\mathcal{R}}^{2W}$ defined as before. In general: $\bar{\mathcal{R}}^F \subseteq \bar{\mathcal{R}}^K \subseteq \bar{\mathcal{R}}^{2W}$. There exists channels such that $\bar{\mathcal{R}}^F \subsetneq \bar{\mathcal{R}}^K$.

*Proof:* The first statement follows directly from the definition of the key generation strategy in Section IV. For the second statement, consider the example of a channel for which:

$$0 < \rho_1 < \frac{h_2 - 1}{h_1} \text{ and } 0 < \rho_2 < \frac{h_1 - 1}{h_2}.$$

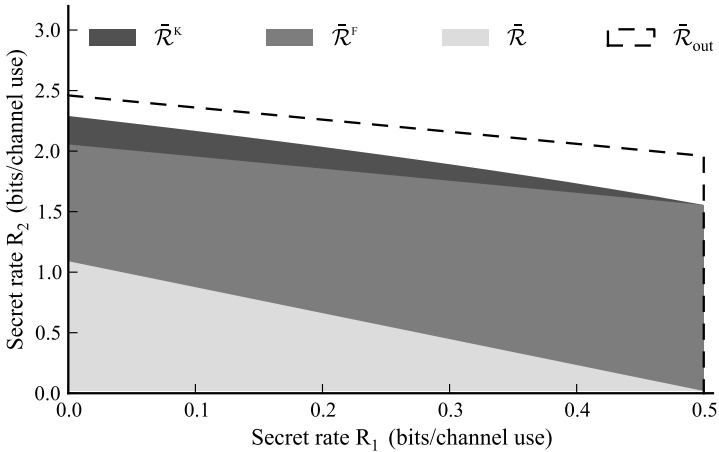

Fig. 4. Region evaluation for $\rho_1 = 1$, $\rho_2 = 100$, $h_1 = 1$, $h_2 = 0.1$, and $g_1 = g_2 = 1$

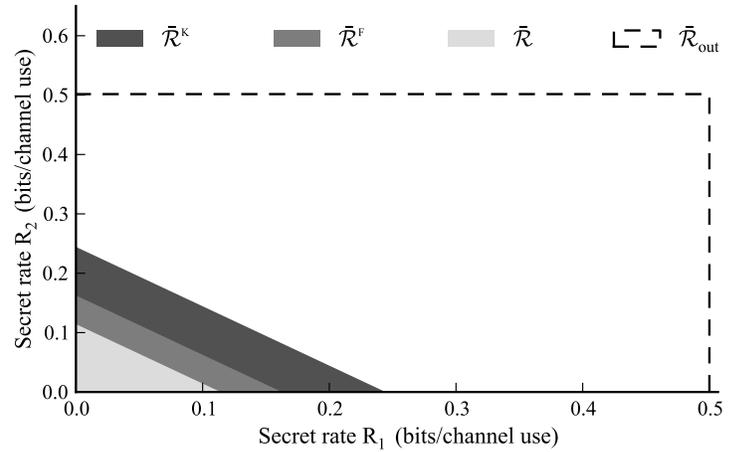

Fig. 5. Region evaluation for $\rho_1 = \rho_2 = 1$, $h_1 = h_2 = 1.5$, $g_1 = g_2 = 1$

In this case $\mathbb{I}(C_1;Y_2|X_2) < \mathbb{I}(C_1;Z)$ and $\mathbb{I}(C_2;Y_1|X_1) < \mathbb{I}(C_2;Z)$, and the region $\bar{\mathcal{R}}^{\text{F}}$ is empty. However, since the power constraints are non-zero, a fraction of the power can be used to induce a DMS while still maintaining a positive auxiliary message rate. By [17, Theorem 3], it is possible to obtain a non-zero key rate and, therefore, a non-zero secure communication rate. ∎

*Remark* Although we only prove $\bar{\mathcal{R}}^{\text{F}} \subsetneq \bar{\mathcal{R}}^{\text{K}}$ by exhibiting a dummy example, our numerical results in the next section clearly show improvements in various cases.

*B. Numerical results*

We first consider a situation in which secret key generation provides little gain because cooperative jamming leaves little room for improvement. Fig. 4 illustrates the region $\bar{\mathcal{R}}^{\text{F}}$ and $\bar{\mathcal{R}}^{\text{K}}$ obtained for $\rho_1 = 1$, $\rho_2 = 100$, $h_1 = 1$, $h_2 = 0.1$, and $g_1 = g_2 = 1$. The upper left corner of the region corresponds to a situation in which Alice uses all her power to jam ($\rho_1^n = \rho_1$) while Bob uses all his power to transmit ($\rho_2^c = \rho_2$). A secret key can be extracted from the DMS induced by Alice; however, since $h_1 = 10$, Eve's source observation is highly correlated to Alice's, which severely limits key rates. The upper right corner of the region corresponds to a situation in which both Alice and Bob use their entire power to transmit (i.e. $\rho_1^c = \rho_1$ and $\rho_2^c = \rho_2$) and, therefore, no DMS is induced. For comparison, we also plot the region $\bar{\mathcal{R}}$ obtained using key generation alone (no cooperative jamming). Although this strategy is clearly suboptimal, note that there exist practical coding schemes for key generation over Gaussian channel; therefore, the region $\bar{\mathcal{R}}$ provides an estimate of rates achievable with current codes. We also plot the best known outer region $\bar{\mathcal{R}}_{\text{out}}$ for the two-way wiretap channel computed by He and Yener [26].

Fig. 5 illustrates the region $\bar{\mathcal{R}}^{\text{F}}$ and $\bar{\mathcal{R}}^{\text{K}}$ obtained for $\rho_1 = \rho_2 = 1$, $h_1 = h_2 = 1.5$, and $g_1 = g_2 = 1$. This corresponds to a situation in which Eve obtains a better observation than either Alice and Bob and in which Alice and Bob have little power available. In such a case, key generation provides a significant improvement. Finally, Fig. 6 illustrates the region $\bar{\mathcal{R}}^{\text{F}}$ and $\bar{\mathcal{R}}^{\text{K}}$ obtained for $\rho_1 = \rho_2 = 0.9$, $h_1 = h_2 = 10$, and $g_1 = g_2 = 1$. In this case $\bar{\mathcal{R}}^{\text{F}} = \varnothing$, but key generation is possible. This case corresponds to the situation used in the proof of Proposition 4.

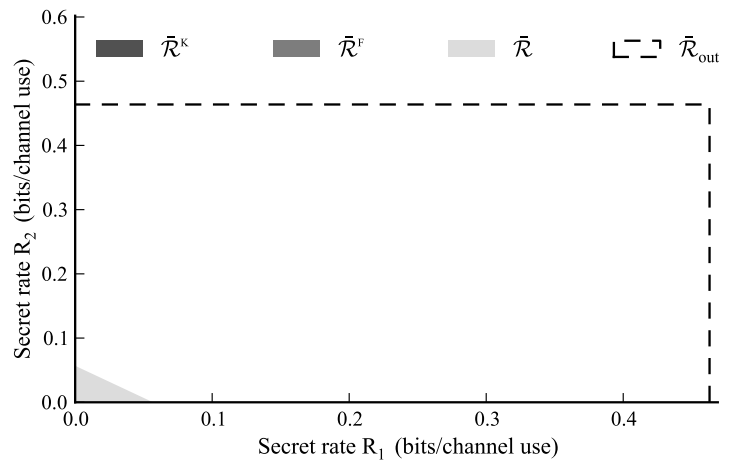

Fig. 6. Region evaluation for $\rho_1 = \rho_2 = 0.9$, $h_1 = h_2 = 10$, $g_1 = g_2 = 1$

## VI. CONCLUSION AND PERSPECTIVES

In this paper, we have developed an achievable region of strongly secure rates based on strategies that partially decouple the use of interference and feedback for secrecy. We have first exploited interference by using a cooperative jamming strategy on top of which we have introduced some basic feedback mechanisms. We have relied on key exchange to transfer secure rate between users and we have performed key generation to distill secret keys from the noise sources induced by cooperative jamming. Numerical results in the Gaussian case have shown significant improvement in some cases.

Our results can naturally be extended in several directions. First, note that the key generation process we considered is based on one-way communication between the two users. There exist sources for which two-way communication is provably better than one-way communication [3]; we can exhibit a wiretap channel based on those sources for which the same result would hold. Second, we can extend the Gaussian two-way wiretap channel to multiple antennas by relying on the results of Khisti and Wornell [27] and Oggier and Hassibi [28] for the MIMO wiretap channel, and Watanabe and Oohama [17] for secret key agreement from vector sources. Third, we could generalize our results to more than two legitimate users. While this poses no major conceptual problem, generalizing Lemma 2 might not be so straightforward. Fourth, the results obtained for the Gaussian case rely on an explicit characterization of the DMS induced by cooperative jamming; it would be interesting to generalize the approach beyond channels with an additive noise structure. Finally, our results generalize to known channels with memory, such as wireless channels with ergodic fading [29], [30], by using the general result of Steinberg on channel resolvability for the multiple-access channel [19].

## APPENDIX A
## PROOF OF LEMMA 1

Let us define: $M_1 \triangleq \lceil 2^{nR_1} \rceil$, $M_2 \triangleq \lceil 2^{nR_2} \rceil$, $M'_1 \triangleq \lceil 2^{nR'_1} \rceil$ and $M'_2 \triangleq \lceil 2^{nR'_2} \rceil$, and let $C$ be the random variable representing a randomly generated code. Let[1] $\epsilon > 0$.

We show that, without loss of generality, we can restrict our attention to the transmission of a single message:

$$\begin{aligned}
\mathbb{E}(\mathrm{P}_e(C)) &= \mathbb{E}\left(\sum_{\substack{\mu_1 \in \mathcal{M}_1 \times \mathcal{M}'_1 \\ \mu_2 \in \mathcal{M}_2 \times \mathcal{M}'_2}} \mathrm{P}_e(C|\mu_1, \mu_2 \text{ sent}) \mathbb{P}(\mu_1, \mu_2 \text{ sent})\right) \\
&= \sum_{\substack{\mu_1 \in \mathcal{M}_1 \times \mathcal{M}'_1 \\ \mu_2 \in \mathcal{M}_2 \times \mathcal{M}'_2}} \mathbb{P}(\mu_1, \mu_2 \text{ sent}) \underbrace{\mathbb{E}_C(\mathrm{P}_e(C|\mu_1, \mu_2 \text{ sent}))}_{\mathbb{E}_C(\mathrm{P}_e(C|\mathbf{1},\mathbf{1} \text{ sent}))} \\
&\stackrel{(a)}{=} \mathbb{E}(\mathrm{P}_e(C|\mathbf{1},\mathbf{1} \text{ sent})) \\
&\stackrel{(b)}{\leqslant} \mathbb{E}(\mathbb{P}(\mu_1 \neq \hat{\mu}_1 | \mathbf{1},\mathbf{1} \text{ sent}, C)) \\
&\quad + \mathbb{E}(\mathbb{P}(\mu_2 \neq \hat{\mu}_2 | \mathbf{1},\mathbf{1} \text{ sent}, C)). \quad (23)
\end{aligned}$$

Equality (a) follows from the symmetry of the random code construction, while inequality (b) is given by the union bound.

Each term can be expressed using the following events:
- $\mathcal{E}(i,j) = \{(X_1^n, C_2^n(i,j), Y_1^n) \in \mathcal{A}_{1,\epsilon}^n\}$
- $\mathcal{F}(i,j) = \{(X_2^n, C_1^n(i,j), Y_2^n) \in \mathcal{A}_{2,\epsilon}^n\}$

Now, we compute:

$$\begin{aligned}
&\mathbb{E}(\mathbb{P}(\mu_2 \neq \hat{\mu}_2 | \mathbf{1},\mathbf{1} \text{ sent}, C)) \\
&= \mathbb{P}\left(\mathcal{E}^c(1,1) \vee \bigvee_{(i,j) \neq \mathbf{1}} \mathcal{E}(i,j)\right) \\
&\leqslant \underbrace{\mathbb{P}(\mathcal{E}^c(1,1))}_{\leqslant \delta(\epsilon)} + \sum_{(i,j) \neq \mathbf{1}} \mathbb{P}(\mathcal{E}(i,j)) \\
&\leqslant \delta(\epsilon) + \sum_{(i,j) \neq \mathbf{1}} \sum_{(x_1^n, c_2^n, y_1^n) \in \mathcal{A}_{1,\epsilon}^n} p_{X_1^n Y_1^n}(x_1^n, c_1^n) p_{C_2^n}(c_2^n) \\
&\leqslant \delta(\epsilon) + \underbrace{M_2 M'_2}_{\leqslant \exp_2(n(R_2 + R'_2 + \delta(n)))} \exp_2(n(\mathbb{H}(X_1 C_2 Y_1) + \epsilon)) \\
&\qquad\qquad \exp_2(-n(\mathbb{H}(C_2) - \epsilon)) \cdot \exp_2(-n(\mathbb{H}(X_1 Y_1) - \epsilon)) \\
&= \delta(\epsilon) + \exp_2(n(R_2 + R'_2 - \mathbb{I}(C_2; Y_1 | X_1) + \delta(n) + 3\epsilon)). \\
&\qquad\qquad\qquad\qquad\qquad\qquad\qquad\qquad\qquad\qquad (24)
\end{aligned}$$

The second inequality follows from the union bound, and the penultimate from the AEP.

If the rate constraints in (6) are satisfied, there exists a $\gamma > 0$ such that $R_2 + R'_2 < \mathbb{I}(Y_1; C_2 | X_1) - \gamma$, and choosing $\epsilon$ such that $\gamma + 3\epsilon > 0$, we obtain:

$$\lim_{n \to \infty} \mathbb{E}_C(\mathbb{P}(\mu_2 \neq \hat{\mu}_2 | \mathbf{1},\mathbf{1} \text{ sent}, C)) \leqslant \delta(\epsilon).$$

By symmetry, we obtain, by a similar reasoning on $\mathcal{F}$:

$$\lim_{n \to \infty} \mathbb{E}_C(\mathbb{P}(\mu_1 \neq \hat{\mu}_1 | \mathbf{1},\mathbf{1} \text{ sent}, C)) \leqslant \delta(\epsilon).$$

Therefore: $\lim_{n \to \infty} \mathbb{E}(\mathrm{P}_e(C)) \leqslant \delta(\epsilon)$.　　**QED**

## APPENDIX B
## PROOF OF LEMMA 2

Let[1] $\epsilon > 0$.

*An upper bound for the leakage using divergence:*

$$\begin{aligned}
\mathrm{L}(\mathcal{C}) &\triangleq \mathbb{I}(Z^n; M_1 M_2 | \mathcal{C}) \qquad (25) \\
&= \mathbb{D}(p_{M_1 M_2 Z^n | \mathcal{C}} \| p_{M_1 M_2 | \mathcal{C}} p_{Z^n | \mathcal{C}}) \\
&= \sum_{m_1=1}^{M_1} \sum_{m_2=1}^{M_2} \sum_{z^n \in \mathcal{Z}^n} p_{M_1 M_2 Z^n | \mathcal{C}}(m_1, m_2, z^n) \\
&\qquad\qquad \log_2 \frac{p_{M_1 M_2 Z^n | \mathcal{C}}(m_1, m_2, z^n)}{p_{M_1 M_2 | \mathcal{C}}(m_1, m_2) p_{Z^n | \mathcal{C}}(z^n)} \\
&= \sum_{m_1=1}^{M_1} \sum_{m_2=1}^{M_2} \sum_{z^n \in \mathcal{Z}^n} p_{Z^n | M_1 M_2, \mathcal{C}}(z^n | m_1, m_2) \\
&\qquad p_{M_1 M_2}(m_1, m_2) \log_2 \frac{p_{Z^n | M_1 M_2, \mathcal{C}}(z^n | m_1, m_2)}{p_{Z^n | \mathcal{C}}(z^n)} \\
&= \frac{1}{M_1 M_2} \sum_{m_1=1}^{M_1} \sum_{m_2=1}^{M_2} \mathbb{D}(p_{Z^n | m_1, m_2, \mathcal{C}} \| p_{Z^n | \mathcal{C}}) \\
&\leqslant \frac{1}{M_1 M_2} \sum_{m_1=1}^{M_1} \sum_{m_2=1}^{M_2} \mathbb{D}(p_{Z^n | m_1, m_2, \mathcal{C}} \| p_{Z^n | \mathcal{C}}) \\
&\qquad\qquad + \mathbb{D}(p_{Z^n | \mathcal{C}} \| p_{Z^n}) \text{ because: } \mathbb{D}(\cdot \| \cdot) \geqslant 0 \\
&= \frac{1}{M_1 M_2} \sum_{m_1=1}^{M_1} \sum_{m_2=1}^{M_2} \mathbb{D}(p_{Z^n | m_1, m_2, \mathcal{C}} \| p_{Z^n}). \qquad (26)
\end{aligned}$$

Taking the expectation and using the symmetry of the random code construction, yields:

$$\begin{aligned}
\mathbb{E}_C(\mathrm{L}(C)) &= \mathbb{I}(Z^n; M_1 M_2 | C) \\
&\leqslant \frac{1}{M_1 M_2} \sum_{m_1=1}^{M_1} \sum_{m_2=1}^{M_2} \mathbb{E}_C\big(\mathbb{D}(p_{Z^n | m_1, m_2, C} \| p_{Z^n})\big) \\
&= \mathbb{E}_C\big(\mathbb{D}(p_{Z^n | 1, 1, C} \| p_{Z^n})\big). \qquad (27)
\end{aligned}$$

---

[1]For convenience we shall add extra conditions on $\epsilon$ as needed.

*An upper bound for the divergence: :*

$$\mathbb{D}(p_{Z^n|1,1,\mathcal{C}}\|p_{Z^n})$$
$$= \sum_{z^n \in \mathcal{Z}^n} p_{Z^n|1,1,\mathcal{C}}(z^n) \log_2 \frac{p_{Z^n|1,1,\mathcal{C}}(z^n)}{p_{Z^n}(z^n)}$$
$$= \sum_{z^n \in \mathcal{Z}^n} p_{Z^n|1,1,\mathcal{C}}(z^n) \log_2 \frac{p_{Z^n|1,1,\mathcal{C}}(z^n)}{p_{Z^n}(z^n)}$$
$$\mathbb{1}\left\{\log_2 \frac{p_{Z^n|1,1,\mathcal{C}}(z^n)}{p_{Z^n}(z^n)} \leqslant \epsilon\right\}$$
$$+ \sum_{z^n \in \mathcal{Z}^n} p_{Z^n|1,1,\mathcal{C}}(z^n) \log_2 \frac{p_{Z^n|1,1,\mathcal{C}}(z^n)}{p_{Z^n}(z^n)}$$
$$\mathbb{1}\left\{\log_2 \frac{p_{Z^n|1,1,\mathcal{C}}(z^n)}{p_{Z^n}(z^n)} > \epsilon\right\}$$
$$\leqslant \epsilon + \sum_{z^n \in \mathcal{Z}^n} p_{Z^n|1,1,\mathcal{C}}(z^n) \log_2 \frac{p_{Z^n|1,1,\mathcal{C}}(z^n)}{p_{Z^n}(z^n)}$$
$$\mathbb{1}\left\{\log_2 \frac{p_{Z^n|1,1,\mathcal{C}}(z^n)}{p_{Z^n}(z^n)} > \epsilon\right\}. \quad (28)$$

By the law of total probability:

$$p_{Z^n}(z^n)$$
$$= \sum_{c_1^n \in \mathcal{C}_1^n} \sum_{c_2^n \in \mathcal{C}_2^n} p_{Z^n|C_1^n C_2^n}(z^n|c_1^n, c_2^n) p_{C_1^n}(c_1^n) p_{C_2^n}(c_2^n)$$
$$\geqslant \sum_{c_1^n \in \mathcal{B}_1(1)} \sum_{c_2^n \in \mathcal{B}_2(1)} p_{Z^n|C_1^n C_2^n}(z^n|c_1^n, c_2^n) p_{C_1^n}(c_1^n) p_{C_2^n}(c_2^n)$$
$$\geqslant \underbrace{M_1' M_2' p_{\min_1}^n p_{\min_2}^n}_{p_{\min}^n} p_{Z^n|1,1,\mathcal{C}}(z^n),$$

where $\mathcal{B}_i(m)$, for $i \in \{1,2\}$, represents the set of codewords associated with message $m$ for user $i$. To obtain the last inequality, the p.d.f. of $C_1$ and $C_2$ must have compact support, which can always be obtained by cropping (to get a compact support) and scaling (to keep an unit area) the p.d.f. If $\check{C}_1$ and $\check{C}_2$ are derived from $C_1$ and $C_2$ with discrete or Gaussian p.d.f., it is possible to get $\mathbb{V}(p_{C_1}, p_{\check{C}_1})$ and $\mathbb{V}(p_{C_2}, p_{\check{C}_2})$ as small as we want. Since $\mathbb{I}(X;Y)$ viewed as a function of $p_X$, with $p_{Y|X}$ fixed, is continuous, the mutual informations involved in (8) are hardly modified by using $\check{C}_1$ and $\check{C}_2$ instead of $C_1$ and $C_2$.

By taking the expectation of (28) and with [19], we have:

$$\mathbb{E}_C(\mathrm{L}(C)) = \mathbb{I}(Z^n; M_1 M_2 | C) \leqslant \epsilon + n \log_2\left(\frac{1}{p_{\min}}\right) J_\epsilon, \quad (29)$$

where $J_\epsilon$ is defined as:

$$J_\epsilon \triangleq$$
$$\sum_{c_1^{(2)} \in \mathcal{C}_1^n} \cdots \sum_{c_1^{(M_1')} \in \mathcal{C}_1^n} \sum_{c_2^{(2)} \in \mathcal{C}_2^n} \cdots \sum_{c_2^{(M_2')} \in \mathcal{C}_2^n}$$
$$p_{C_1^n}\left(c_1^{(2)}\right) \cdots p_{C_1^n}\left(c_1^{(M_1')}\right) p_{C_2^n}\left(c_2^{(2)}\right) \cdots p_{C_2^n}\left(c_2^{(M_2')}\right)$$
$$\times \sum_{c_1^{(1)} \in \mathcal{C}_1^n} \sum_{c_2^{(1)} \in \mathcal{C}_2^n} \sum_{z^n \in \mathcal{Z}^n} p_{C_1^n C_2^n Z^n}\left(c_1^{(1)}, c_2^{(1)}, z^n\right)$$
$$\times \mathbb{1}\left\{\frac{1}{M_1' M_2'} \exp_2\left(i_{C_1^n C_2^n; Z^n}\left(c_1^{(1)}, c_2^{(1)}; z^n\right)\right)\right.$$
$$+ \frac{1}{M_1' M_2'} \sum_{(i,j) \neq (1,1)} \exp_2\left(i_{C_1^n C_2^n; Z^n}\left(c_1^{(i)}, c_2^{(j)}; z^n\right)\right)$$
$$\left. > 1 + 4\epsilon \right\}.$$

*An upper bound for $J_\epsilon$:* We remind the following assumption from (8):
$$\begin{cases} R_1' + R_2' > \mathbb{I}(C_1 C_2; Z) \\ R_1' > \mathbb{I}(C_1; Z) \\ R_2' > \mathbb{I}(C_2; Z) \end{cases}$$

The quantity $J_\epsilon$ can be upper bounded as (see [19]):

$$J_\epsilon \leqslant J^{(1)} + J^{(2)} + J^{(3)} + J^{(4)}, \quad (30)$$

where $J^{(1)}$, $J^{(2)}$, $J^{(3)}$, and $J^{(4)}$ are defined as:

$$J^{(1)} = \mathbb{P}\left(\frac{1}{M_1' M_2'} \exp_2\left(i_{C_1^n C_2^n; Z^n}\left(C_1^{(1)} C_2^{(1)}; Z^n\right)\right) > \epsilon\right)$$
$$J^{(2)} = \mathbb{P}\left(\frac{1}{M_1' M_2'} \sum_{i=1}^n \sum_{j=1}^n \exp_2\left(i_{C_1^n C_2^n; Z^n}\left(C_1^{(i)} C_2^{(j)}; Z^n\right)\right)\right.$$
$$\left. > 1 + \epsilon\right)$$
$$J^{(3)} = \mathbb{P}\left(\frac{1}{M_1' M_2'} \sum_{i=1}^n \exp_2\left(i_{C_1^n C_2^n; Z^n}\left(C_1^{(i)} C_2^{(1)}; Z^n\right)\right) > \epsilon\right)$$
$$J^{(4)} = \mathbb{P}\left(\frac{1}{M_1' M_2'} \sum_{j=1}^n \exp_2\left(i_{C_1^n C_2^n; Z^n}\left(C_1^{(1)} C_2^{(j)}; Z^n\right)\right) > \epsilon\right)$$

Now, we study each term individually:

- First, $J^{(1)}$ can be upper-bounded as follows:
$$J^{(1)} \leqslant \mathbb{P}(\exp_2(I(C_1^n C_2^n; Z^n)) >$$
$$\exp_2(-n\epsilon + nR_1' + nR_2'))$$
$$= \mathbb{P}\left(\frac{1}{n} I(C_1^n C_2^n; Z^n) > -\epsilon + R_1' + R_2'\right)$$
$$= \mathbb{P}\left(\frac{1}{n} \sum_{i=1}^n I(C_{1,i} C_{2,i}; Z_i) > R_1' + R_2' - \epsilon\right)$$

since $R_1' + R_2' > \mathbb{I}(C_1 C_2; Z)$, there exists $\gamma > 0$ such that $R_1' + R_2' > \mathbb{I}(C_1 C_2; Z) + \gamma$. If we choose $\epsilon < \gamma$ to guarantee $R'_1 + R'_2 - \epsilon > \mathbb{I}(C_1 C_2; Z) + \underbrace{\gamma - \epsilon}_{>0}$ and to apply the Chernoff bound, we have:

$$\exists \alpha_{\gamma-\epsilon} > 0, \ J^{(1)} \leqslant e^{-\alpha_{\gamma-\epsilon} n}.$$

- With a similar reasoning, we can upper bound $J^{(2)}$, $J^{(3)}$, and $J^{(4)}$ by quantities going exponentially to zero under the given conditions.

*Conclusion:* $J_\epsilon$ goes exponentially to zero as $n$ goes to infinity. Therefore, $\lim_{n\to\infty} \mathbb{E}(\mathrm{L}(C)) \leqslant \delta(\epsilon)$. **QED**

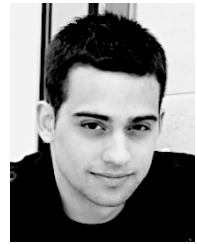

**Alexandre J. Pierrot** (S'11) received the Diplôme d'Ingénieur from Supélec, Gif-sur-Yvette, France, in 2011 and the M.Sc. degree in electrical and computer engineering from the Georgia Institute of Technology, Atlanta, GA, in 2011, where he is currently pursuing the Ph.D. degree.

He is working in the communication architecture research group (Arcom) at Georgia Tech-Lorraine, Metz, France. His research interests include physical layer security, digital communications, and digital signal processing.

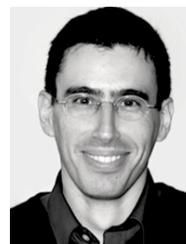

**Matthieu R. Bloch** (S'07–M'08) received the Diplôme d'Ingénieur from Supélec, Gif-sur-Yvette, France, the M.Sc. degree in electrical and computer engineering from the Georgia Institute of Technology, Atlanta, GA, in 2003, the Ph.D. degree in engineering from the Université de Franche-Comté, Besancon, France, in 2006, and the Ph.D. degree in electrical engineering from the Georgia Institute of Technology in 2008.

In 2008-2009, he was a postdoctoral research associate at the University of Notre Dame, South Bend, IN, USA. Since July 2009, he has been on the faculty of the School of Electrical and Computer Engineering at the Georgia Institute of Technology, where he is currently an Assistant Professor based at the Georgia Tech-Lorraine campus, Metz, France. His research interests are in the areas of information theory, error-control coding, wireless communications, and cryptography.

Dr. Bloch is the recipient of the 2011 ComSoc and Information Theory Society Joint Paper Award, and the co-author of Physical-Layer Security: *From Information Theory to Security Engineering*, which will be published by Cambridge University Press in July 2011.